\documentclass[conference,10pt]{IEEEtran}
\IEEEoverridecommandlockouts
\usepackage{amssymb}
\usepackage{amsmath,amsthm,amsfonts}\usepackage[english]{babel}
\usepackage{dsfont}

\usepackage{xspace}
\usepackage{cases}
\usepackage[table,svgnames]{xcolor}
\usepackage{graphicx}
\usepackage[font=footnotesize]{caption}
\usepackage[font=footnotesize]{subcaption}
\usepackage{epstopdf}
\usepackage{upgreek}
\usepackage{tikz}
\usetikzlibrary{shapes.geometric, arrows}
\usepackage{booktabs}
\usepackage{array}
\usepackage{blindtext}
\usepackage{multirow}
\usepackage{bbold}
\usepackage{soul}
\usepackage{enumerate}
\usepackage{blindtext}
\usepackage{siunitx}
\usepackage{pgf}
\usepackage{lipsum}
\usepackage[
colorlinks=true,
linkcolor =red,
urlcolor  = blue,
citecolor =blue
]{hyperref}
\usepackage[acronym,hyperfirst=true]{glossaries}
\usepackage{enumitem}
\usepackage{setspace}
\hypersetup{
  colorlinks=true,
  citecolor=Blue,
  linkcolor=Red,
  urlcolor=Blue}
\setitemize[1]{itemsep=4.5pt,partopsep=1pt,parsep=\parskip,topsep=4.5pt}
\usepackage{algorithm,algorithmic}

\setkeys{glslink}{hyper=false}

\newtheorem{coro}{Corollary}

\newacronym{opf}{OPF}{optimal power flow}
\newacronym{qp}{\textsc{qp}}{quadratic program}
\newacronym{nlp}{\textsc{nlp}}{nonlinear programming}
\newacronym{rapidpf}{rapid\textsc{pf}}{rapid prototyping for distributed Power Flow}
\newacronym{admm}{\textsc{admm}}{Alternating Direction Method of Multipliers}
\newacronym{aladin}{\textsc{aladin}}{Augmented Lagrangian based Alternating Direction Inexact Newton method}
\newacronym{ocd}{\textsc{ocd}}{Optimality Condition Decomposition}
\newacronym{app}{\textsc{app}}{Auxiliary Problem Principle}
\newacronym{sqp}{\textsc{sqp}}{Sequential Quadratic Programming}
\newacronym{milp}{\textsc{milp}}{Mixed-Integer Linear Programming}
\usepackage{titlesec}
\titlespacing*{\section}{0pt}{*0.8}{*0.8}
\titlespacing{\subsection}{0pt}{*0.8}{*0.8}
\renewcommand{\thesubsubsection}{\arabic{subsubsection}}

\definecolor{RB}{rgb}{.1,.4,.9}
\definecolor{VH}{rgb}{0.,.8,.4}
\definecolor{revised}{rgb}{.2,.8,.1}

\newacronym{tso}{TSO}{transmission system operator}
\newacronym{dso}{DSO}{distribution system operator}
\newacronym{dsos}{DSOs}{distribution system operators}
\newacronym{der}{DER}{distributed energy resource}
\newacronym{ders}{DERs}{distributed energy resources}
\newacronym{pcc}{PCC}{point of common coupling}
\newacronym{bim}{BIM}{bus injection model}
\newacronym{ess}{ESS}{energy storage systems}
\newacronym{soc}{SOC}{state of charge}
\newacronym{itd}{ITD}{integrated transmission-distribution}

\makeglossaries
\usepackage{cleveref}
\crefname{figure}{Fig.}{Figs.} 
\Crefname{figure}{Fig.}{Figs.} 
\crefname{equation}{Eq.}{Eqs.} 
\Crefname{equation}{Eq.}{Eqs.} 
\usepackage{subcaption}
\usepackage{makecell}

\newcommand{\abs}[1]{\left|#1\right|}

\ifCLASSINFOpdf
\else
\fi
\def\BibTeX{{\rm B\kern-.05em{\sc i\kern-.025em b}\kern-.08em
    T\kern-.1667em\lower.7ex\hbox{E}\kern-.125emX}}
\begin{document}

\title{{Error Accumulation} using Linearized Models for Aggregating Flexibility in Distribution Systems
}

\author{\IEEEauthorblockN{Yanlin Jiang\IEEEauthorrefmark{2},
Xinliang Dai\IEEEauthorrefmark{1},
Frederik Zahn\IEEEauthorrefmark{2},
Yi Guo\IEEEauthorrefmark{3},
and
Veit Hagenmeyer\IEEEauthorrefmark{2}}
\IEEEauthorblockA{\IEEEauthorrefmark{2}Institute for Automation and Applied Informatics,
Karlsruhe Institute of Technology, Germany\\
\IEEEauthorrefmark{1}Andlinger Center for Energy and the Environment, Princeton University, USA\\
\IEEEauthorrefmark{3}Urban Energy Systems Laboratory, Empa, Switzerland.\\
Email: \{yanlin.jiang, frederik.zahn, veit.hagenmeyer\}@kit.edu; xinliang.dai@princeton.edu; yi.guo@empa.ch}\\
\thanks{\IEEEauthorrefmark{1}Corresponding Author: X.~Dai (xinliang.dai@princeton.edu).}
}

\maketitle

\begin{abstract}
This paper investigates flexibility aggregation approaches based on linear models. We begin by examining the theoretical foundations of linear AC power flow, two variants of {so-called} DC power flow, and the LinDistFlow model, along with their underlying assumptions. The discussion covers key system details, including network topology, voltage constraints, and line losses. Simulations are conducted on the KIT Campus Nord network with real demand and solar data. Results show that, in the absence of negative losses, line losses are generally underestimated by linear models. Furthermore, line losses errors tend to accumulate both at the point of common coupling (PCC) and over extended time horizons.
\end{abstract}


\section{Introduction}
In recent decades, the integration of \acrfull{ders} into power systems has grown rapidly, with most of these resources being decentralized and connected to distribution networks. This shift requires closer TSO-DSO collaboration to ensure efficient and secure \acrshort{ders}' mangement~\cite{itd2020review,kerscher2022key}. While a centralized management approach could, in theory, optimize the entire power system, it is impractical due to computational limits, privacy concerns, and high communication demands~\cite{molzahn2017survey,patari2021distributed,dai2025largescale}.

As an alternative, flexibility aggregation provides a non-iterative distributed coordination framework: each \acrshort{dso} independently determines its implicit feasible set of net power exchange with the \acrshort{tso}, within which safe operation of the distribution system can be ensured~\cite{chen2020aggregate}. However, accurately characterizing this feasible set is challenging because of the nonlinearity in AC power systems, making the dispatch problem NP-hard even for radial networks~\cite{lehmann2015ac}.

Several linear models have been developed to simplify the power flow problem. Among them, the {DC power flow model}~\cite{stott2009dc} is one of the most widely used. However, this model is lossless and primarily designed for transmission systems, where voltage and angle differences are minor, and branch reactances dominate over resistances; these assumptions are typically not valid in distribution networks. Another lossless approach is the LinDistFlow model, a simplification of the DistFlow model that neglects power losses~\cite{lopez2021quickflex}. LinDistFlow is formulated explicitly for radial distribution networks, reflecting their typical topological characteristics. To incorporate power losses into linear models, two main approaches are typically used: either the power losses themselves are linearized, or the full AC power flow equations are linearized directly. For the former, the enhanced DC model is proposed~\cite{yang2017linearized}.\footnote{The model introduced in~\cite{yang2017linearized} is referred to as the linearized optimal power flow model. To distinguish it from other linear models discussed in this paper, we refer to it as the enhanced DC power flow model.} This model adopts the common assumptions of small voltage magnitude and angle differences to eliminate trigonometric functions, and power losses are explicitly linearized. By contrast, the linearized AC power flow model is derived without relying on such assumptions. Instead, once a power flow solution is obtained, the system Jacobian is used to capture the relationship between system variables and demand~\cite{contreras2021congestion}.


Leveraging the convexity of the linear models, different flexibility aggregation methods have been introduced to aggregate flexibility. These include outer approximations~\cite{zhang2023coordination}, inner approximations~\cite{wang2021aggregate,wang2025non,wen2023improvedDER}, boundary detection techniques~\cite{chen2020aggregate,chen2021leveraginga}, data-driven approaches~\cite{lyu2025data}, and projection-based methods such as Fourier-Motzkin elimination~\cite{wen2022tdder}.  Building on this, \cite{dai2024realtime} employs a temporal decomposition strategy, allowing the feasible region at each time step to be computed independently and in parallel.

However, both the omission of line losses and the use of linearized models can lead to significant inaccuracies in flexibility aggregation. As shown in \cite{lopez2021quickflex,jiang2025enhanced}, flexibility derived from the LinDistFlow model often deviates from results based on the exact AC model. Additionally, linearization of losses may even result in physically unrealistic outcomes, such as negative losses. To mitigate this, \cite{yang2017linearized} proposes adding a penalty term to the objective function. However, such techniques are difficult to incorporate into flexibility aggregation frameworks.

The approach in~\cite{jiang2025enhanced} aggregates flexibility at a specific time point. We build on these results to reveal the impact of system losses on the multiperiod scheduling problem. To this end, we first provide a comprehensive review of linear models used for flexibility aggregation. A real distribution network is then used to evaluate different models for flexibility aggregation and solve the corresponding scheduling problem. Our results show that, in open-loop scheduling scenarios, underestimated power losses accumulate over time. This leads to over-discharging of the \acrfull{ess}, highlighting the importance of incorporating real-time feedback or correction mechanisms into the scheduling process.

\section{AC Power Flow Models}~\label{sec::pf}

This section presents two variants of AC power flow equations for power systems, both derived from the complex power flow formulation.

\subsection{Preliminaries}

A power system can be represented as a graph $\mathcal{S} = (\mathcal{N}, \mathcal{L})$, where $\mathcal{N}$ denotes the set of buses and $\mathcal{L}$ the set of branches. The subsets $\mathcal{L}^f$ and $\mathcal{L}^t$ correspond to the from-side and to-side branches, respectively.

The AC power flow equations describe the fundamental relationships among key quantities in the power system, namely current $I$, voltage $V$, admittance $Y$, and power $S$. Three fundamental electrical principles underpin these equations. First, Kirchhoff’s Current Law (KCL) states that the sum of currents flowing into and out of a bus must be balanced. Mathematically, this is expressed as 
\begin{equation} 
    \qquad \qquad I_i = \sum_{(i,j)\in\mathcal{L}} I_{ij},\qquad \qquad \forall i \in\mathcal{N}
\end{equation}
where $I_i$ denotes the current injection from devices, including load and generator, at bus $i$, and $I_{ij}$ is the current along branch $(i, j)$. Second, Ohm’s Law relates the current along a branch to the voltage difference between its terminals: 
\begin{equation}\label{eq::ohm}
    \qquad \qquad I_{ij} = Y_{ij} (V_i-V_j),\qquad \qquad  \forall (i,j) \in\mathcal{L}
\end{equation}
with $Y_{ij}$ being the admittance of the branch between buses $i$ and $j$. For more on the admittance matrix $Y$, see \cite[Sec.~4]{frank2016introduction}. Third, the definition of AC power specifies the complex power flow 
\begin{equation}\label{eq::pf}
    \qquad \qquad S_{ij} = Y_{i} I^*_{ij},\qquad \qquad  \forall (i,j) \in\mathcal{L}
\end{equation}
where $V_i$ is the voltage at bus $i$ and $I^*_{ij}$ is the complex conjugate of the branch current.

By combining these properties, we obtain the standard AC power flow model. The power balance at each bus $i$ is
\begin{subequations}\label{eq::pf::complex}
    \begin{align}
        S_i =& \sum_{(i,j)\in\mathcal{L}} S_{ij},&\forall i\in\;&\mathcal{N}\label{eq::nodalBalance::complex}
    \end{align}
    and the power flow along branch $(i, j)$ is 
    \begin{align}
        S_{ij}=&\quad Y^*_{ij} V_i V_i^* - Y_{ij}^*V_iV_J^*,\;&\forall (i,j)\in\;&\mathcal{L},\label{eq::powerflow::complex}
    \end{align}
\end{subequations}
Collectively, these nonlinear equations~\eqref{eq::pf::complex} form the foundation for many power system applications, governing nodal power balance and power flows across the network.

\subsection{AC Power Flow Equations in Polar Coordinates}
When voltages are represented in polar coordinates,
\begin{align}\label{eq::polar::voltage}
    V_i = v_i e^{\,\textbf{j}\theta_i},\quad \forall i\in\mathcal{N}, 
\end{align}
the AC power flow equations~\eqref{eq::pf::complex}  can be reformulated in terms of these polar coordinates:
\begin{subequations}\label{eq::polar::pf}
    \begin{align}
    & p_i =v_i \sum_{j\in\mathcal{N}} v_j \left( g_{ij} \cos\theta_{ij} + b_{ij} \sin\theta_{ij} \right), &\forall i&\in\mathcal{N},\label{eq::pfe::polar1}\\
    &q_i =v_i \sum_{j\in\mathcal{N}} v_j \left( g_{ij} \sin\theta_{ij} - b_{ij} \cos\theta_{ij} \right),&\forall i&\in\mathcal{N},\label{eq::pfe::polar2}
\end{align}
\end{subequations}
where $v_i$ and $\theta_i$ are the magnitude and angle of the complex voltage $V_i$,  $g_{ij}$ and $b_{ij}$ are the real and imaginary parts of the admittance matrix $Y_{ij}$, and $\theta_{ij} = \theta_i - \theta_j$ is the phase angle difference between buses $i$ and $j$. 
Note that the power flow balance
\begin{subequations}\label{eq::pf::polar}
    \begin{align}
        p_{ij}&=+ v_i^2g_{ij}-v_i v_j\left(g_{ij}\cos \theta_{ij}+b_{ij}\sin \theta_{ij}\right), \\
        q_{ij}&=-v_i^2b_{ij} - v_i v_j\left(g_{ij}\sin \theta_{ij}-b_{ij}\cos \theta_{ij}\right),
    \end{align}
\end{subequations}
for all {branches $\forall(i,j)\in\mathcal{L}$ are implicitlty} included in the nodal balance equations~\eqref{eq::polar::pf}.

\subsection{DistFlow Model}

For radial distribution systems, the DistFlow model~\cite{baran1989optimal1,baran1989optimal2} is frequently used. Let $z_{ij} = r_{ij}+\mathbf{j}x_{ij}$ be the complex impedance on the line. Then Ohm’s law~\eqref{eq::ohm} and complex power flow~\eqref{eq::pf} can be rewritten as
\begin{subequations}\label{eq::distflow::complex}
    \begin{align}\label{eq::distflow::complex::1}
    V_i = V_j + z_{ij} \frac{S^*_{ij}}{V_i^*},\qquad\forall (i,j)\in \mathcal{L}^f
    \end{align}
and the complex nodal balance~\eqref{eq::nodalBalance::complex} becomes
    \begin{align}\label{eq::distflow::complex::2}
        S_{i} & =  \sum_{(i,j)\in\mathcal{L}^f}\hspace{-8pt} S_{ij}-\hspace{-8pt}\sum_{(i,j)\in\mathcal{L}^t} \hspace{-5pt}(S_{ij}-z_{ij} \abs{I_{ij}}^2),&\forall i\in\mathcal{N}
    \end{align}
\end{subequations}
By the magnitude squared of~\eqref{eq::distflow::complex::1}, and the decomposition of the complex nodal balance~\eqref{eq::distflow::complex::2} into real and imaginary parts, we can obtain the DistFlow equations:
\begin{subequations}\label{eq::distflow}
    \begin{align}
p_i &= \hspace{-6pt}\sum_{(i,j)\in\mathcal{L}^f}\hspace{-6pt} P_{ij}
      - \sum_{(i,j)\in\mathcal{L}^t} \hspace{-6pt}\bigl(P_{ij} - r_{ij}\ell_{ij}\bigr),
      &\forall j \in\mathcal{N} \\
q_i &= \hspace{-6pt}\sum_{(i,j)\in\mathcal{L}^f} \hspace{-6pt}Q_{ij}
      - \sum_{(i,j)\in\mathcal{L}^t}  \hspace{-6pt}\bigl(Q_{ij} - x_{ij}\ell_{ij}\bigr),
      &\forall j \in\mathcal{N} \\
u_i &= u_j +    2\bigl(r_{ij}P_{ij} + x_{ij}Q_{ij}\bigr)\notag\\
      &\qquad\qquad\qquad\qquad - (r_{ij}^2 + x_{ij}^2)\ell_{ij},\hspace{-5pt}
      &\forall (i,j)\in \mathcal{L}^f\\
\ell_{ij} &= \frac{P_{ij}^2 + Q_{ij}^2}{u_i},
      &\forall (i,j)\in \mathcal{L}^f 
\end{align}
\end{subequations}
with squared magnitudes of the current $\ell_{ij} = \abs{I_{ij}}^2$ and the voltage $u_i = \abs{V_i}^2$.

The DistFlow model~\eqref{eq::distflow} is derived by eliminating explicit angle information, and thus serves as an angle-relaxed version of the optimal sizing of capacitors power flow model~\eqref{eq::pf::complex}. This angle relaxation is exact when the network is radial.
\begin{coro}[\textbf{Theorem 4~\cite{farivar2013branch}}]
    When $\mathcal{G}$ is a tree, the DistFlow model~\eqref{eq::distflow} is equivalent to the complex power flow model~\eqref{eq::pf::complex}.
\end{coro}
This result confirms that, under radial topology, both models~\eqref{eq::polar::pf}~\eqref{eq::distflow} describe the same set of feasible operating points for the power system.

\section{Aggregation using Linear Models}~\label{sec::aggregation}
This section provides a brief overview of four linear models used for flexibility aggregation, all derived from the nonlinear AC power flow models discussed above. We also briefly describe the methodology for flexibility aggregation. A summary of these linear models is presented in Table~\ref{tab::comparison}.

\subsection{LinDistFlow}

For radial distribution systems, one may start from the DistFlow~\eqref{eq::distflow} and drop the quadratic losses terms $r_{ij}\ell_{ij}$ and $x_{ij}\ell_{ij}$, obtaining
\begin{subequations}\label{eq::lindistflow}
    \begin{align}
p_i &= \hspace{-6pt}\sum_{(i,j)\in\mathcal{L}^f}\hspace{-6pt} P_{ij}
      - \sum_{(i,j)\in\mathcal{L}^t} \hspace{-6pt}P_{ij},
      &\forall j \in\mathcal{N} \\
q_i &= \hspace{-6pt}\sum_{(i,j)\in\mathcal{L}^f} \hspace{-6pt}Q_{ij}
      - \sum_{(i,j)\in\mathcal{L}^t}  \hspace{-6pt}Q_{ij} ,
      &\forall j \in\mathcal{N} \\
u_i &= u_j +    2\bigl(r_{ij}P_{ij} + x_{ij}Q_{ij}\bigr) &\forall (i,j)\in \mathcal{L}^f
\end{align}
\end{subequations}
Because \eqref{eq::lindistflow} ignores those quadratic loss terms, it is most accurate when system losses are modest, but it can underestimate voltages when DER output is high. In flexibility aggregation, ignoring losses means that errors may accumulate at the PCC~\cite{jiang2025enhanced}.

\subsection{Classic DC power flow}
The classic DC power flow model arises from the AC power flow equations~\eqref{eq::polar::pf} under certain simplifying assumptions: line resistances $r_{ij}$ are negligible, voltage magnitudes are close to unity, and angle differences are small. That is,
\begin{subequations}\label{eq::DC::approximation}
    \begin{align}
        &\sin{\theta_{ij}} \approx \theta_{ij},\;\cos{\theta_{ij}} \approx 1,\;&\forall (i,j)\in\mathcal{L}\\
        &v_i \approx 1,\;&\forall i\in\mathcal{N}
    \end{align}
\end{subequations}
Under these approximations,  the AC power flow equations~\eqref{eq::polar::pf} can be reduced to:
\begin{subequations}\label{eq::dc::pf}
    \begin{align}
        p_i &= \sum_{(i,j)\in\mathcal{L}} P_{ij}&\forall i\in\mathcal{N}\\
        P_{ij} & = b_{ij} \theta_{ij} & \forall (i,j)\in\mathcal{L}
        \end{align}
\end{subequations}

Reactive injections and voltage magnitudes are ignored, yielding a lossless, active-power-only model. This approach is accurate in high-voltage transmission systems, but less suitable for low-voltage distribution systems with significant $R/X$ ratios.

\subsection{Enhanced DC Power Flow}
To increase the fidelity of the DC power flow model, \cite{yang2017linearized} proposed enhancements that consider line losses, voltage differences, and reactive power. They improve the standard approximations~\eqref{eq::DC::approximation} as follows:
\begin{subequations}
    \begin{align}\label{eq::DC::approximation::enhance}
        &\sin{\theta_{ij}} \approx \theta_{ij},\;\cos{\theta_{ij}} \approx 1-\frac{\theta_{ij}^2}{2},\;&\forall (i,j)\in\mathcal{L}\\
        &v_i v_j \theta_{ij} \approx \theta_{ij},v_i v_j \theta_{ij}^2 \approx \theta_{ij}^2,\;&\forall i\in\mathcal{N}
    \end{align}
\end{subequations}
{This yields an enhanced version of the DC power flow:}
\begin{subequations}\label{eq::dc::pf::enhanced}
    \begin{align}
        p_i &= \sum_{(i,j)\in\mathcal{L}} P_{ij},\quad
        q_i = \sum_{(i,j)\in\mathcal{L}} Q_{ij}&\forall i\in\mathcal{N}\\
        p_{ij} &= +\,g_{ij}\frac{(u_i^2-u_j^2)}{2} - b_{ij}\theta_{ij} +P^\text{losses}_{ij},
        &\forall (i,j)\in\mathcal{L}\\
        q_{ij} &=  -\,b_{ij}\frac{(u_i^2-u_j^2)}{2} - g_{ij}\theta_{ij} +Q^\text{losses}_{ij},&\forall(i,j)\in\mathcal{L}
        \end{align}        
\end{subequations}
with the squared voltage magnitudes $u_i = \abs{V_i}^2$ and $u_{ij} = (\abs{V_i}-\abs{V_j})^2$, where the nonlinear losses term
\begin{subequations}\label{eq::dc::losses}
    \begin{align}
        P^\text{losses}_{ij} = g_{ij}\frac{u_{ij}+\theta_{ij}}{2},\,Q^\text{losses}_{ij} = -b_{ij}\frac{u_{ij}+\theta_{ij}}{2}
    \end{align}        
\end{subequations}
are approximated linearly based on an operation point; see~\cite[Appendix~A]{yang2017linearized} for details about linear approxiamtion. However, these linear models may yield negative losses under certain conditions~\cite {hobbs2008improved}.

\subsection{Linear Approximation of AC Power Flow}

Let $x=\{v_i,\theta_i,p_i,q_i\}_{i\in\mathcal{N}}$ collect voltage magnitude and angle, and active and reactive power injections at all buses. Let $F(x)$ denote the nonlinear AC power flow equations~\eqref{eq::polar::pf}. The linearized AC power flow model is given by the first-order Taylor expansion of $F(\cdot)$ at a nominal operating point $x_0$:

\begin{equation}
    \label{eq::polar::pf::linear}
    \frac{\partial F}{\partial x}  \cdot \Delta x = 0.
\end{equation}
with $\Delta x = x-x_0$. The inverse Jacobian $[\frac{\partial F}{\partial x}]^{-1}$ is readily available from the Newton–Raphson power-flow solver and yields high-fidelity voltage sensitivities as long as operating conditions remain within the linear neighbourhood of $x_0$. Far from that point, however, the approximation may violate voltage limits or lead to the negative losses~\cite{hobbs2008improved}.

\begin{table*}[htbp!]
    \centering
    \caption{Linear Models for Flexibility Aggregation}
    \footnotesize
    \begin{tabular}{@{}l|ccccc|ccc@{}}
        \toprule
        & \multicolumn{5}{c}{\textbf{Model Detail}} & \multicolumn{2}{c}{\textbf{Simulation Results}}\\
        & Topology & Voltage Magnitude & Voltage Angle &   Reactive Power&Line Loss& Negative Loss & Optimality\\
        \midrule
        LinDistFlow&       radial &squared&  -  &standard& - &- &+ \\
        Classic DC PF&            meshed &standard& -  & -&-&-  & + \\
        {Enhanced} DC PF&    meshed &squared& standard&  standard &{linearized} &NaN& + + \\
        Linearized AC PF & meshed &standard& standard&  standard &{linearized}  &NaN& + + +\\
        \bottomrule
    \end{tabular}
    \label{tab::comparison}
\end{table*}

\subsection{Flexibility Aggregation with Linearized Models}
The dispatch problem for a distribution system can be expressed as the following optimization:
\begin{subequations}\label{eq::flexibility::linear}
    \begin{align}
        \min_{x,y} \quad & f(x) + g(y)\label{eq::flexibility::linear::obj}\\
        \text{s.t.} \quad & Ax+By - c=0\label{eq::flexibility::linear::c1}\\
        & (x,\;y) \in \mathcal{X}\times \mathcal{Y}\label{eq::flexibility::linear::c2}
    \end{align}
\end{subequations}
With close sets 
$$\mathcal{X}=\{x\,\mid\,\underline{x}\leq x \leq \overline{x}\},\text{ and }\mathcal{Y}=\{x\,\mid\,\underline{y}\leq y \leq \overline{y}\}.$$
{where} $x$ represents the coupling variables, such as the net power exchange at the \acrshort{pcc}, while $y$ contains all other variables local to the distribution system. Constraint~\eqref{eq::flexibility::linear::c1} enforces the power flow constraints, and~\eqref{eq::flexibility::linear::c2} captures the operational limits. Together, they define a polyhedral feasible set.

Flexibility aggregation focuses on characterizing the implicit set of coupling variables $x$ for which there exists a feasible $y$ that satisfies all constraints~\eqref{eq::flexibility::linear::c1}~\eqref{eq::flexibility::linear::c2}. This defines the implicit feasible region,
$$\mathcal{X}^\text{imp}=\{x \in\mathcal{X}\,\mid\, Ax+By-c = 0,\;y\in\mathcal{Y}\}.$$
Then, the optimization problem can be reformulated as
\begin{subequations}\label{eq::flexibility::aggregate}
    \begin{align}
        \min_{x} \quad & f(x) + g(\hat{y})\\
        \text{s.t.} \quad & x\in\mathcal{X}^\text{imp}
    \end{align}
\end{subequations}
This approach allows flexibility to be aggregated and described as an implicit feasible set with respect to the coupling variables, greatly simplifying coordination with the transmission system.

\section{Simulation}
\label{sec::simulation}
This section compares scheduling results obtained using aggregated flexibility models based on linear power flow approximations. In the simulation, the active power at the PCC and the total SOC of all ESSs serve as the coupling variables $x$, and $y$ indicate the rest state variables. The linear models used for aggregation include LinDistFlow~\eqref{eq::lindistflow}, classic DC power flow~\eqref{eq::dc::pf}, enhanced DC power flow~\eqref{eq::dc::pf::enhanced}, and linearized AC power flow~\eqref{eq::polar::pf::linear}, all of which are reformulated into the compact structure shown in \eqref{eq::flexibility::linear}. We apply the flexibility aggregation method introduced in \cite{jiang2025enhanced} to obtain a set of simplified linear constraints that represent the coupling at the PCC. Scheduling results solved under these coupling constraints are compared to the full AC power flow model (\eqref{eq::polar::pf}), which serves as the benchmark. The simulation follows a day-ahead scheduling framework, where DSOs plan multiperiod dispatches based on forecast data for the upcoming day.
\subsection{Setting}

\begin{figure}[htbp]
    \centering
    \includegraphics[width=0.95\linewidth]{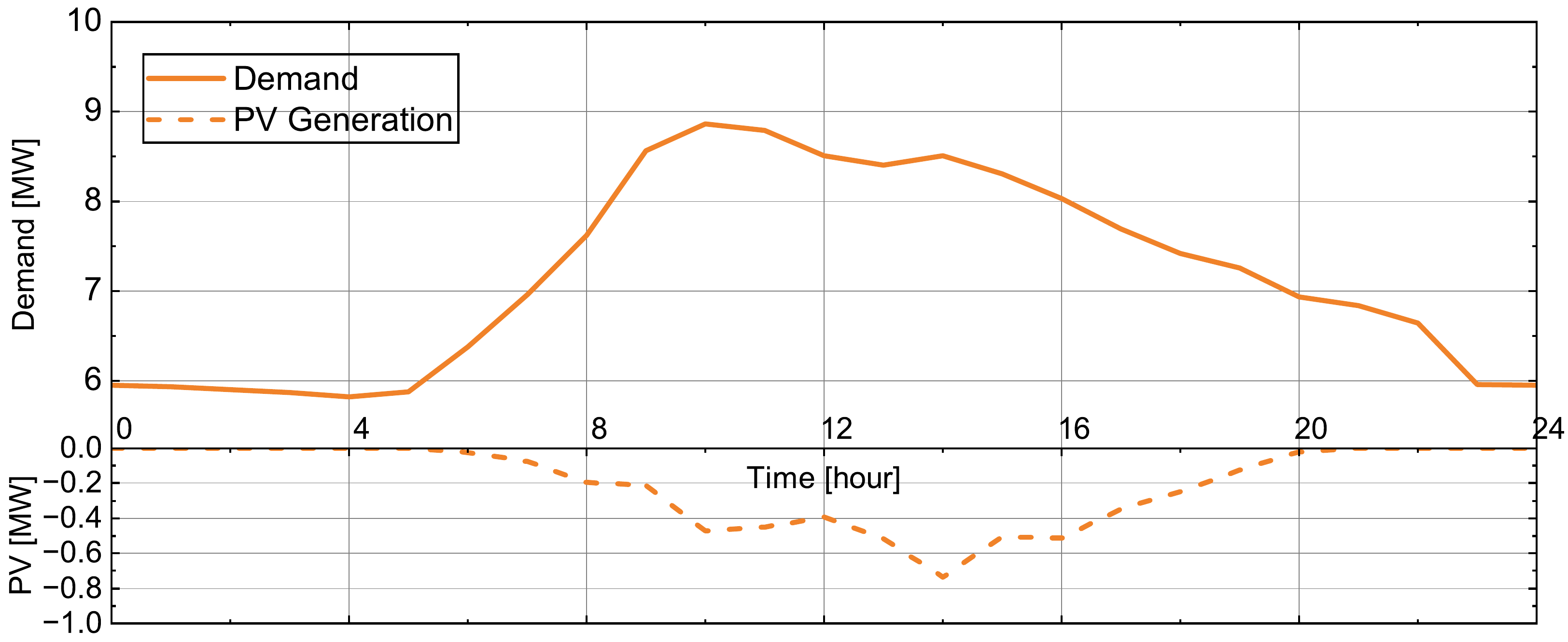}
    \label{fig::demand}
    \caption{Temporal Profiles of prosumption for a Typical Day}
    \label{fig::prosumption}
\end{figure}

We use a real distribution network from the North Campus of the Karlsruhe Institute of Technology (KIT) as the test case. The network operates in a radial topology and includes 40 buses. A PV system is installed at bus 33. To evaluate the impact of flexibility on scheduling, we add three \acrshort{ess}s at buses 13, 26, and 39. Each battery is sized to charge or discharge fully in two hours at maximum power. For simplicity in scheduling, the initial and final \acrfull{soc} is set to 50\%, and the (dis-) /charging efficiency is assumed to be 100\%, maintaining the linearity of the model. The scheduling tests use historical load data with hourly resolution, covering both demand and PV generation. For Day-Ahead scheduling, we use representative medium prosumption profiles on workdays as the forecast, shown in Figure~\ref{fig::prosumption}. 


{Specifically, both the enhanced DC model and the linearization of the AC power flow model require a predefined base point. To obtain this, the average prosumption over the entire year is used as input to the AC power flow problem, whose solution is then used for linearization.}



\subsection{Day-Ahead Scheduling}

In the Day-Ahead stage, scheduling is carried out based on the given predicted prosumption. The objective function is to minimize the quadratic total system cost. As a benchmark, we first solve the scheduling problem using the full AC power flow model. Next, we apply the flexibility aggregation method to different linear models, based on the provided prosumption at each time step, the feasible flexibility is aggregated as a power-energy envelope at the \acrshort{pcc} and used as constraints in the scheduling process. 

Once the scheduling problem leveraging aggregated flexibility is solved, the practical coordination between transmission and distribution networks is considered. This coordination requires the transmission system to deliver the power requested by the DSO, assuming a fixed injection at the \acrshort{pcc}. As a result, a post-verification step is necessary to ensure that the scheduled power injection $P_\text{pcc}$ is feasible for the distribution network. To perform this check, we substitute the scheduled $P_\text{pcc}$ into the exact AC power flow model and analyze the resulting \acrshort{soc} over time. 

\begin{figure}[htbp]
    \centering
    \begin{subfigure}{\linewidth}
        \centering
        \caption{Active power exchanges at PCC}
        \vspace{0.2em}
        \includegraphics[width=0.95\linewidth]{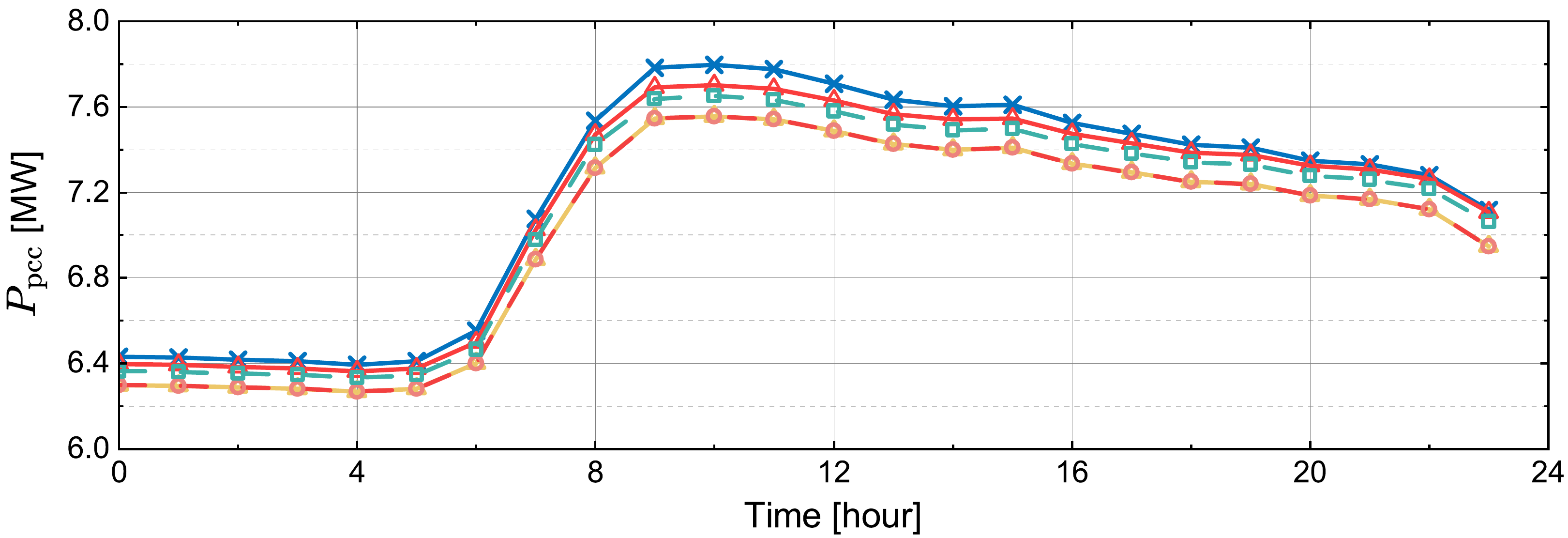}
        \label{fig::Ppcc}
    \end{subfigure}

    \begin{subfigure}{\linewidth}
        \centering
        \caption{State of Charge after verification using AC model}
        \vspace{0.2em}
        \includegraphics[width=0.95\linewidth]{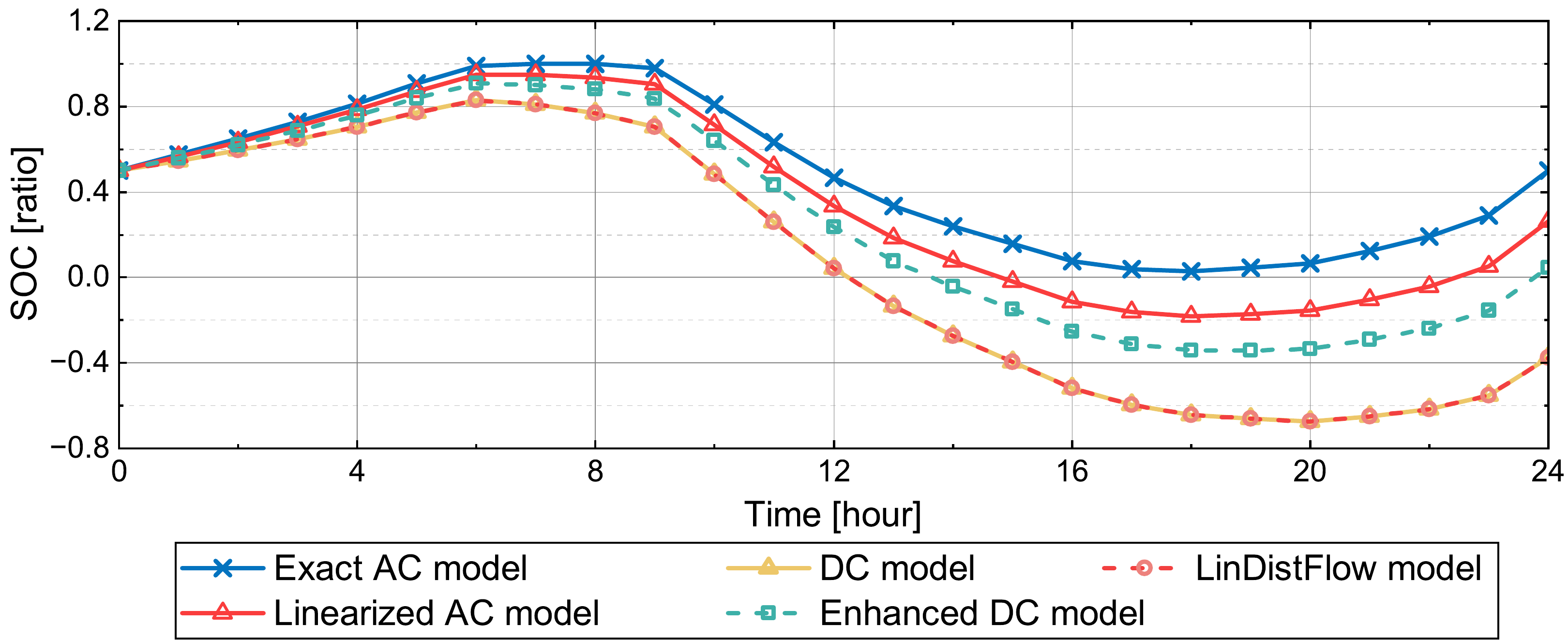}
        \label{fig::SOC}
    \end{subfigure}
    \vspace{0em}
    \caption{Day-Ahead Scheduling with Post-AC-Verification}
    \label{fig::scheduling}
\end{figure}
Figure~\ref{fig::Ppcc} shows the scheduled power exchanged at the \acrshort{pcc}, based on aggregated flexibility models, which represents the power requested by the DSO and delivered by the TSO. The Figure~\ref{fig::SOC} illustrates the resulting SOC after substituting the scheduled $P_\text{pcc}$ values into the exact AC power flow model for verification.
 
Compared to the reference trajectory obtained from the exact AC power flow model, the linear models demonstrate similar scheduling behaviors. However, the DC and LinDistFlow models, which neglect power losses, consistently schedule lower $P_\text{pcc}$ values than the reference. As a result, the total energy consumption (including system losses) is underestimated. This leads to an over-discharge of the \acrshort{ess} after 12:00 a.m., causing its \acrshort{soc} to drop below zero and fail to reach the target final value of 50\%. This discrepancy is confirmed through the models' inherent problem, {which neglects system losses, resulting in cumulative errors within the \acrshort{ess}.} Linear models that include linearized losses perform slightly better, but still encounter similar challenges. These arise from the fact that linearized losses cannot accurately capture variations across periods. Moreover, their accuracy is sensitive to the choice of the linearization base point. Note that negative losses did not occur in our simulation. Overall, the performance of different linear models is listed in the right part of Table~\ref{tab::comparison}. 




\section{Conclusion \& Outlook}
\label{sec::conclusion}
The present paper evaluates the use of various linear power flow models for flexibility aggregation and scheduling of \acrshort{ess}s in distribution networks. Simulation results show that inaccurate modeling of losses can lead to infeasible scheduling outcomes, particularly for energy storage systems, where underestimation of consumption may result in over-discharge. {Using linear line-loss accounting models, such as the enhanced DC and linearized AC power flow, their performance still results in over-discharging}, which is attributed to the differences between varying demand and the fixed demand for linearization. Future work will explore real-time correction strategies, such as receding horizon control, to compensate for the accumulated losses within the \acrshort{ess}s.

\bibliographystyle{ieeetr}
\bibliography{transLib}

\begin{thebibliography}{10}

\bibitem{itd2020review}
A.~G. Givisiez, K.~Petrou, and L.~F. Ochoa, ``A review on {TSO}-{DSO} coordination models and solution techniques,'' {\em Electric Power Systems Research}, vol.~189, p.~106659, 2020.

\bibitem{kerscher2022key}
S.~Kerscher and P.~Arboleya, ``The key role of aggregators in the energy transition under the latest european regulatory framework,'' {\em International Journal of Electrical Power \& Energy Systems}, vol.~134, p.~107361, 2022.

\bibitem{molzahn2017survey}
D.~K. Molzahn, F.~D{\"o}rfler, H.~Sandberg, S.~H. Low, S.~Chakrabarti, R.~Baldick, and J.~Lavaei, ``A survey of distributed optimization and control algorithms for electric power systems,'' {\em EEE Transactions on Smart Grid}, vol.~8, no.~6, pp.~2941--2962, 2017.

\bibitem{patari2021distributed}
N.~Patari, V.~Venkataramanan, A.~Srivastava, D.~K. Molzahn, N.~Li, and A.~Annaswamy, ``Distributed optimization in distribution systems: Use cases, limitations, and research needs,'' {\em IEEE Transactions on Power Systems}, vol.~37, no.~5, pp.~3469--3481, 2021.

\bibitem{dai2025largescale}
X.~Dai, Y.~Jiang, Y.~Guo, C.~N. Jones, M.~Diehl, and V.~Hagenmeyer, ``Distributed ac optimal power flow: A scalable solution for large-scale problems,'' {\em arXiv preprint arXiv:2503.24086}, 2025.

\bibitem{chen2020aggregate}
X.~Chen, E.~Dall'Anese, C.~Zhao, and N.~Li, ``Aggregate {{Power Flexibility}} in {{Unbalanced Distribution Systems}},'' {\em IEEE Transactions on Smart Grid}, vol.~11, pp.~258--269, Jan. 2020.

\bibitem{lehmann2015ac}
K.~Lehmann, A.~Grastien, and P.~Van~Hentenryck, ``{AC}-feasibility on tree networks is {NP}-{H}ard,'' {\em IEEE Transactions on Power Systems}, vol.~31, no.~1, pp.~798--801, 2015.

\bibitem{stott2009dc}
B.~Stott, J.~Jardim, and O.~Alsa{\c{c}}, ``Dc power flow revisited,'' {\em IEEE Transactions on Power Systems}, vol.~24, no.~3, pp.~1290--1300, 2009.

\bibitem{lopez2021quickflex}
L.~Lopez, A.~{Gonzalez-Castellanos}, D.~Pozo, M.~Roozbehani, and M.~Dahleh, ``{{QuickFlex}}: A {{Fast Algorithm}} for {{Flexible Region Construction}} for the {{TSO-DSO Coordination}},'' in {\em 2021 {{International Conference}} on {{Smart Energy Systems}} and {{Technologies}} ({{SEST}})}, (Vaasa, Finland), pp.~1--6, IEEE, Sept. 2021.

\bibitem{yang2017linearized}
Z.~Yang, H.~Zhong, A.~Bose, T.~Zheng, Q.~Xia, and C.~Kang, ``A linearized opf model with reactive power and voltage magnitude: A pathway to improve the mw-only dc opf,'' {\em IEEE Transactions on Power Systems}, vol.~33, no.~2, pp.~1734--1745, 2017.

\bibitem{contreras2021congestion}
D.~A. Contreras, S.~M{\"u}ller, and K.~Rudion, ``Congestion {{Management Using Aggregated Flexibility}} at the {{TSO-DSO Interface}},'' in {\em 2021 {{IEEE Madrid PowerTech}}}, pp.~1--6, June 2021.

\bibitem{zhang2023coordination}
T.~Zhang, J.~Wang, H.~Wang, J.~Ruiyang, G.~Li, and M.~Zhou, ``On the {{Coordination}} of {{Transmission-Distribution Grids}}: {{A Dynamic Feasible Region Method}},'' {\em IEEE Transactions on Power Systems}, vol.~38, no.~2, pp.~1857--1868, 2023.

\bibitem{wang2021aggregate}
S.~Wang and W.~Wu, ``Aggregate {{Flexibility}} of {{Virtual Power Plants With Temporal Coupling Constraints}},'' {\em IEEE Transactions on Smart Grid}, vol.~12, pp.~5043--5051, Nov. 2021.

\bibitem{wang2025non}
S.~Wang, C.~Feng, and F.~You, ``Non-iterative coordination of interconnected power grids via dimension-decomposition-based flexibility aggregation,'' {\em arXiv preprint arXiv:2502.07226}, 2025.

\bibitem{wen2023improvedDER}
Y.~Wen, Z.~Hu, J.~He, and Y.~Guo, ``Improved inner approximation for aggregating power flexibility in active distribution networks and its applications,'' {\em EEE Transactions on Smart Grid}, vol.~early access, 2023.

\bibitem{chen2021leveraginga}
X.~Chen and N.~Li, ``Leveraging {{Two-Stage Adaptive Robust Optimization}} for {{Power Flexibility Aggregation}},'' {\em IEEE Transactions on Smart Grid}, vol.~12, pp.~3954--3965, Sept. 2021.

\bibitem{lyu2025data}
R.~Lyu, H.~Guo, G.~Strbac, and C.~Kang, ``Data-driven dimension reduction for industrial load modeling using inverse optimization,'' {\em IEEE Transactions on Smart Grid}, 2025.

\bibitem{wen2022tdder}
Y.~Wen, Z.~Hu, and L.~Liu, ``Aggregate temporally coupled power flexibility of ders considering distribution system security constraints,'' {\em IEEE Transactions on Power Systems}, vol.~38, no.~4, pp.~3884--3896, 2023.

\bibitem{dai2024realtime}
X.~Dai, Y.~Guo, Y.~Jiang, C.~N. Jones, G.~Hug, and V.~Hagenmeyer, ``Real-time coordination of integrated transmission and distribution systems: {{Flexibility}} modeling and distributed {{NMPC}} scheduling,'' {\em Electric Power Systems Research}, vol.~234, p.~110627, 2024.

\bibitem{jiang2025enhanced}
Y.~Jiang, X.~Dai, F.~Zahn, and V.~Hagenmeyer, ``Enhanced flexibility aggregation using {LinDistFlow} model with loss compensation,'' {\em arXiv preprint arXiv:2505.01715}, 2025.

\bibitem{frank2016introduction}
S.~Frank and S.~Rebennack, ``An introduction to optimal power flow: Theory, formulation, and examples,'' {\em IIE Transactions}, vol.~48, no.~12, pp.~1172--1197, 2016.

\bibitem{baran1989optimal1}
M.~E. Baran and F.~F. Wu, ``Optimal capacitor placement on radial distribution systems,'' {\em IEEE Transactions on Power Delivery}, vol.~4, no.~1, pp.~725--734, 1989.

\bibitem{baran1989optimal2}
M.~Baran and F.~F. Wu, ``Optimal sizing of capacitors placed on a radial distribution system,'' {\em IEEE Transactions on Power Delivery}, vol.~4, no.~1, pp.~735--743, 1989.

\bibitem{farivar2013branch}
M.~Farivar and S.~H. Low, ``Branch flow model: Relaxations and convexification ({P}arts {I}, {II}),'' {\em IEEE Transactions on Power Systems}, vol.~28, no.~3, pp.~2554--2564, 2013.

\bibitem{hobbs2008improved}
B.~F. Hobbs, G.~Drayton, E.~B. Fisher, and W.~Lise, ``Improved transmission representations in oligopolistic market models: quadratic losses, phase shifters, and dc lines,'' {\em IEEE Transactions on Power Systems}, vol.~23, no.~3, pp.~1018--1029, 2008.

\end{thebibliography}

\end{document}